\documentclass[preprint,12pt]{elsarticle}

\usepackage{amssymb}
\usepackage{amsthm}
\usepackage{mathtools}

\newcommand{\R}{\mathbb{R}}

\newcommand{\Q}{\mathbb{Q}}

\newcommand{\N}{\mathbb{N}}
\newcommand{\C}{\mathbb{C}}
\newcommand{\Spec}{\textnormal{Spec}}

\newcommand{\tr}{\textnormal{tr}}
\newcommand{\Gal}{\textnormal{Gal}}
\newcommand{\col}{\textnormal{col}}

\newtheorem{thm}{Theorem}

\newtheorem{lem}[thm]{Lemma}

\journal{Discrete Mathematics}

\begin{document}

\begin{frontmatter}


\author{Julien Sorci\cortext[cor1]{}}
\ead{julien.sorci@quantinuum.com}
%

\title{Average Mixing in Quantum Walks of Reversible Markov Chains}



\affiliation{organization={Quantinuum},
            addressline={Terrington House 13-15 Hills Road}, 
            city={Cambridge},
            postcode={CB2 1NL}, 
            state={},
            country={United Kingdom}}

\begin{abstract}
The Szegedy quantum walk is a discrete time quantum walk model which defines a quantum analogue of any Markov chain. The long-term behavior of the quantum walk can be encoded in a matrix called the \textit{average mixing matrix}, whose columns give the limiting probability distribution of the walk given an initial state. We define a version of the average mixing matrix of the Szegedy quantum walk which allows us to more readily compare the limiting behavior to that of the chain it quantizes. We prove a formula for our mixing matrix in terms of the spectral decomposition of the Markov chain and show a relationship with the mixing matrix of a continuous quantum walk on the chain. In particular, we prove that average uniform  mixing in the continuous walk implies average uniform mixing in the Szegedy walk. We conclude by giving examples of Markov chains of arbitrarily large size which admit average uniform mixing in both the continuous and Szegedy quantum walk.
\end{abstract}



\begin{keyword}
Continuous Quantum Walk \sep Szegedy Quantum Walk \sep Markov Chain \sep Mixing Matrix



\end{keyword}

\end{frontmatter}



\section{Introduction}

There are many important probability distributions which arise as the stationary distribution of a Markov chain. For example, certain Markov chains with uniform stationary distribution have been used to approximately count objects which are difficult to enumerate. The algorithm generally works by defining a Markov chain on the set of objects to be counted. An initial state which is easy to prepare is chosen, and transitions at each time step occur by randomly modifying the current state according to some fixed transition rules. When this Markov chain is irreducible and aperiodic with uniform stationary distribution the state after sufficiently many time steps is a uniformly random sample which may be used to approximate the size of the state space. Concrete applications include approximately counting proper colorings and independent sets in graphs \cite{J1995, DG2000}, approximating the permanent of a $01$-matrix \cite{JS1989}, and approximately counting the feasible solutions of certain knapsack problems \cite{MS2004}. For more background on the use of Markov chains for approximate counting, see \cite{JS1989-2}. 

For this reason it is interesting to ask if a quantized version of a Markov chain can allow us to sample from its stationary distribution more quickly. A natural quantization is through a \textit{quantum walk}, which is a quantum analogue of a random walk. In its most general form, a quantum walk is defined by a unitary transition operator that respects the topology of the random walk, and can be either discrete or continuous-time. The model we study here is a discrete quantum walk called the \textit{Szegedy quantum walk}, which quantizes any Markov chain through a product of two non-commuting reflections, and requires no additional information about the Markov chain aside from the transition probabilities \cite{S2004}. As with a Markov chain, each step of a quantum walk induces a probability distribution by considering the modulus-squared of the amplitudes of the basis states, so we might hope that these distributions will allow us to sample from the stationary distribution of the chain. However, there are two difficulties which arise. First, due to the unitarity of the quantum walk, the sequence of probability distributions it induces will not converge in general, which contrasts with the existence of a stationary distribution in an irreducible and aperiodic Markov chain. Second, the Szegedy quantum walk model is a quantum walk defined on a larger state space than that of the Markov chain, and so their respective distributions are not immediately comparable. This is necessitated by the fact that not every digraph admits unitary dynamics on its vertices \cite{S2003}.

A primary objective of this paper is to provide a framework for understanding the limiting behavior of the Szegedy quantum walk which makes it easily comparable to the stationary distribution of the chain it quantizes. The framework that we propose rigorously quantifies the likelihood of the quantum walker landing on a particular site of the Markov chain in the long-run, given that it was initialized in a state that is concentrated on another site. The main object of study is a matrix called the \textit{average mixing matrix}, whose columns encode the limiting distribution of the quantum walk for the various initial starting sites. The study of average distributions and average mixing matrices in quantum walks is not new. For continuous quantum walks on graphs, the average distribution over the vertices of an Abelian circulant graph was studied in \cite{AABEHLT2007} and shown never to be uniform. Following this, Godsil defined an average mixing matrix over the vertices of a graph in \cite{G2013}; some spectral properties of the matrix were proved and it was explicitly computed for several families of graphs. It was then studied for trees in \cite{GGS2017}, and a general analysis on its diagonal entries was carried out in \cite{GGS2019}. For discrete quantum walks on graphs, Aharonov et al. defined an average distribution over the vertices of the graph conditioned on an arbitrary initial state \cite{AAKV2001}. Godsil and Zhan generalized this to an average mixing matrix over the arcs of an arbitrary graph and determined conditions for average uniform mixing over the arcs \cite{GZ2019}.

In the case of a quantum walk on a Markov chain there appears to be much less known. Richter defined an average mixing matrix for a continuous quantum walk on a Markov chain and used it to present a quantum sampling algorithm \cite{R2007}. To our knowledge, no average mixing matrix for the Szegedy quantum walk has been proposed and rigorously analyzed in the literature which gives the limiting probability distribution over the vertices of the chain given a starting state concentrated on a vertex. We follow previous work and consider a time-averaged limiting probability distribution defined by the Szegedy quantum walk. These distributions are always well-defined and can be computed from the spectral idempotents of the quantum walk's transition matrix. As these distributions are defined over the directed arcs of the chain's underlying digraph, we arrive at a distribution over the vertices by summing over the probabilities of its outgoing arcs. Moreover, we condition the quantum walk on a particular coherent encoding of the vertices of the graph, so that our average mixing matrix reads out the probability of landing on a particular vertex given that our initial state is concentrated on another one. We provide a detailed analysis of our average mixing matrix and show a relationship between it and the average mixing matrix of the continuous quantum walk. Lastly, we consider when our proposed limiting distribution in the Szegedy quantum walk is uniform over the vertices, which we refer to as \textit{average uniform mixing}. We show that average uniform mixing in the continuous quantum walk implies average uniform mixing in the Szegedy quantum walk, and give spectral conditions for the former to occur. We additionally construct an infinite family of symmetric Markov chains whose continuous quantum walk and Szegedy quantum walk are both average uniform mixing. 

The outline of the paper is as follows. In Section~\ref{szegedy}, we review the basic definitions of a Markov chain and the Szegedy quantum walk. In addition, we review the spectral decomposition of the quantum walk transition operator. In Section~\ref{average}, we define our proposed average mixing matrix for the Szegedy quantum walk and prove a formula for it in terms of the spectral idempotents of the chain. In Section~\ref{continuouswalks}, we give necessary and sufficient conditions for the continuous quantum walk to be average uniform mixing and show that average uniform mixing for a continuous quantum walk implies average uniform mixing in the Szegedy quantum walk. In Section~\ref{properties}, we prove some basic properties about the average mixing matrix. Lastly, we construct a family of symmetric Markov chains of arbitrarily large size which admit average uniform mixing in both the continuous quantum walk and Szegedy quantum walk in Section~\ref{uniformmixingexamples}.

\section{The Szegedy Quantum Walk}\label{szegedy}

In this section we review some of the basic theory of Markov chains, along with the Szegedy quantum walk. In general, a discrete Markov chain on a finite state space $\Omega$ of cardinality $n$ is defined by a row-stochastic transition matrix, which is a $n \times n$ matrix $P = (p(x,y))_{x, y \in \Omega}$ whose rows and columns are labeled by the elements of $\Omega$, and such that $p(x,y)$ is the probability of transitioning from $x$ to $y$. If $v \in \R^n$ is the state at time $t=0$, then the state at time $t$ is given by $vP^t$. A state $v \in \R^n$ is said to be \textit{stationary} if $vP=v$, that is, $v$ is a left eigenvector of $P$ with eigenvalue $1$. We can view $P$ as a weighted adjacency matrix of a digraph $X$ whose vertex set is $\Omega$ and $(x,y) \in \Omega \times \Omega$ is a directed arc of $X$ if and only if $p(x,y)$ is nonzero. We will refer to $X$ as the digraph of the chain. We will say that $P$ is \textit{irreducible} if $X$ is strongly connected, and say $P$ is \textit{aperiodic} if for all $x \in \Omega$ we have $\gcd\{t \geq 1: (P^t)_{xx}\} =1$. When $P$ is irreducible and aperiodic there is a unique stationary distribution, which we denote by $\pi$. We will say that a Markov chain $P$ with stationary distribution $\pi$ is \textit{reversible} if it satisfies the equations
\begin{eqnarray*}
	\pi(x) p(x,y) = \pi(y) p(y,x),
\end{eqnarray*}
for all $x, y \in \Omega$, which are referred to as the \textit{detailed-balance equations}. A Markov chain is said to be \textit{symmetric} if $P$ is a symmetric matrix. A symmetric Markov chain is reversible and has uniform stationary distribution.

We define the \textit{discriminant} of the matrix $P$ to be the $n \times n$ matrix $D$ with $(x,y)$ entry equal to $\sqrt{p(x,y)p(y,x)}$. Note that $D$ is a real symmetric matrix, and hence is diagonalizable. Moreover, when the Markov chain $P$ is reversible with stationary distribution $\pi$, then a simple computation using the detailed-balance equations shows that $P$ and $D$ are similar matrices related by
\begin{eqnarray}\label{dsimilarity}
	D = D_\pi P D_\pi^{-1},
\end{eqnarray}
where $D_\pi$ is the $n \times n$ diagonal matrix whose $(x,x)$ entry is $\sqrt{\pi(x)}$. 

In general $P$ is a real matrix acting on the vector space $\R^n$, so that the evolution of the chain occurs on the vertices of its digraph $X$. In contrast to this, the Szegedy quantum walk associated to $P$ is a quantum walk defined by a unitary matrix acting on $\C^n \otimes \C^n$. This vector space has the standard basis $\{\mathbf{e}_x \otimes \mathbf{e}_y : x,y \in \Omega \}$,  and we interpret the state $\mathbf{e}_x \otimes \mathbf{e}_y$ as the quantum walker lying on the arc $(x,y)$. The transition matrix for the Szegedy quantum walk is defined as follows. First, for each vertex $x \in \Omega$ define the vector 
\[ \phi_x := \sum_{y \in \Omega} \sqrt{p(x,y)}\mathbf{e}_x \otimes \mathbf{e}_y, \] 
and let $S$ be the matrix whose columns are the vectors $\phi_x$, for $x \in \Omega$. Second, we define the \textit{arc-reversal} matrix, $R$, by $R (\bold{e}_x \otimes \bold{e}_y) = \bold{e}_y \otimes \bold{e}_x$ for all $x,y \in \Omega$. Then the Szegedy quantum walk is the quantum walk defined by the unitary transition matrix $U:= R(2SS^*-I)$, where $*$ denotes the conjugate transpose. We note that the definition we give here is slightly different than the definition in Szegedy's original paper \cite{S2004}. However, the square of the transition matrix defined here is equal to the transition operator defined by Szegedy, and so the two walks are equivalent.

An important feature of a Hermitian or unitary matrix which we will make use of is its spectral decomposition. If $M$ is Hermitian or unitary with distinct eigenvalues $\lambda_1,...,\lambda_m$, then $M$ can be expressed as
\begin{eqnarray}\label{generalspecdecomp}
	M = \sum_{r=1}^m \lambda_r E_r,
\end{eqnarray}
where $E_r$ is the projection onto the $\lambda_r$-eigenspace of $M$. The projections $E_1,...,E_m$ are referred to as the \textit{spectral idempotents} of $M$ and satisfy the following properties:
\begin{enumerate}[(i)]
	\item $E_r E_s = 0$ for all $r \neq s$. 
	\item $E_r^2 = E_r$ for all $r$.
	\item $\sum_{r=1}^mE_r = I$.
	\item $M E_r = \lambda_r E_r$ for all $r$.
\end{enumerate}
The expression in (\ref{generalspecdecomp}) is referred to as the \textit{spectral decomposition} of $M$.  

We review the spectral decomposition of the transition matrix $U$, which is well-known. First note that $U$ is a product of two reflections, since $R^2 = I = (2SS^* - I)^2$. The matrix $2SS^* - I$ is a reflection about the column space of $S$, which we denote by $\col(S)$. As $R$ is also a reflection, then it can be written as $R = 2MM^* - I$ for some matrix $M$ with orthonormal columns. The following lemma can be read off from \cite[Theorem 1]{S2004}. For a similar analysis, we refer the reader to \cite[Theorem 3.3]{Chan_2023}. 


\begin{lem}
Let $U$ be the transition matrix for the Szegedy quantum walk of an arbitrary Markov chain $P$ with discriminant $D$. For each eigenvalue $\lambda$ of $D$ let $E_\lambda$ denote the projection onto the $\lambda$-eigenspace of $D$, and write $\lambda = \cos\theta_\lambda$. Then
\begin{enumerate}[(i)]
	\item The $1$-eigenspace of $U$ is the direct sum
\begin{eqnarray*}
	\big(\col(M) \cap \col(S)\big) \oplus \big(\ker(M^*) \cap \ker(S^*)\big),
\end{eqnarray*}
and the projection onto $\col(M) \cap \col(S)$ is $F_1:=SE_1S^*$.
	\item The $(-1)$-eigenspace of $U$ is the direct sum
\begin{eqnarray*}
	\big(\col(M) \cap \ker(S^*)\big) \oplus \big(\ker(M^*) \cap \col(S)\big),
\end{eqnarray*}
and the projection onto $\ker(M^*) \cap \col(S)$ is $F_{-1}:=SE_{-1}S^*$.
	\item The non-real eigenvalues of $U$ are given by $e^{\pm i \theta_\lambda}$ for $\lambda \neq \pm 1$. The projections onto the $e^{i \theta_\lambda}$ and $e^{- i \theta_\lambda}$-eigenspaces of $U$ are
\begin{eqnarray}\label{plusidem}
	F_\lambda^+:=\frac{1}{2\sin^2(\theta_\lambda)}(S-e^{i\theta_\lambda}RS)E_\lambda(S-e^{i\theta_\lambda}RS)^*,
\end{eqnarray}
and
\begin{eqnarray}\label{minusidem}
	F_\lambda^-:=\frac{1}{2\sin^2(\theta_\lambda)}(S-e^{-i\theta_\lambda}RS)E_\lambda(S-e^{-i\theta_\lambda}RS)^*,
\end{eqnarray}
respectively. 
\end{enumerate}
\end{lem}

As a consequence of these results, if $P$ is any Markov chain then the spectral decomposition of the transition matrix $U$ restricted to the subspace $\col(S)$ is given by
\begin{eqnarray}\label{specdecomp}
	1 \cdot F_1 + \sum_{\lambda \neq \pm 1} \Big( e^{i \theta_\lambda} F_\lambda^+ + e^{-i\theta_\lambda}F_\lambda^-\Big) + (-1) \cdot F_{-1},
\end{eqnarray}
where the sum is over the eigenvalues of $D$ which are not equal to $\pm 1$.

\section{The Average Mixing Matrix}\label{average}

In general, given a unit vector $v \in \C^n$ and time $t \in \N$, the unitarity of $U$ implies that the vector $U^tv$ also has unit magnitude. Therefore $U^tv \circ \overline{U^tv}$ defines a probability distribution, where $\circ$ denotes the entrywise product and $\overline{M}$ denotes the complex conjugate of a matrix or vector $M$. We might ask how the behavior of the sequence of distributions
\begin{eqnarray*}
\Big( U^tv \circ \overline{U^tv} \Big)_{t = 0}^\infty
\end{eqnarray*}
compares to the corresponding sequence given by $P$. However, as shown by Aharonov et al. \cite{AAKV2001} this sequence does not converge in general unless $U$ is the identity matrix. To obtain a well-defined limiting distribution from the quantum walk we may instead consider the time-averaged distribution,
\begin{eqnarray*}
	\lim_{T \rightarrow \infty}\frac{1}{T}\sum_{t=0}^{T-1} \big(U^t v \circ \overline{U^tv}\big),
\end{eqnarray*}
which is sometimes referred to as the Cesaro limit. This defines a probability distribution over the directed arcs of the underlying digraph, and quantifies how concentrated the quantum walk is on any given arc on average. To more easily compare the limiting behavior of the quantum walk to the stationary distribution of the chain, we will define a related time-averaged distribution over the vertices of the digraph. Since the columns of the matrix $S$ are states which are concentrated around the elements of $\Omega$, then to make the notion of initializing the walk by a state concentrated on an element of $\Omega$ we will assume that the initial state lies in $\col(S)$.

We define the \textit{average mixing matrix} of the Szegedy quantum walk to be the $n \times n$ matrix $\widehat{M}$ with $(x,y)$ entry given by
\begin{eqnarray*}
	\widehat{M}_{xy}:= \lim_{T \rightarrow \infty}\frac{1}{T}\sum_{t=0}^{T-1} \sum_{z \in \Omega} (\bold{e}_x \otimes \bold{e}_z)^*  (U^tS) \circ (\overline{U^tS})\bold{e}_y
\end{eqnarray*}
for $x,y \in \Omega$. Here, the $(x,y)$ entry of $\widehat{M}$ can be interpreted as the probability of landing on the vertex $x$ given the initial state $S \bold{e}_y$, where we view the probability of being on the vertex $x$ as the sum of the probabilities over its outgoing arcs. As noted by Aharonov et al. in \cite{AAKV2001}, we can sample from this average distribution by choosing $T$ sufficiently large, uniformly sampling an integer $t$ from the set $\{0,1,..,T-1\}$, and measuring the quantum state $U^tS\bold{e}_y$. 

After a simple translation of \cite[Theorem 3.4]{AAKV2001} into our notation, we can express $\widehat{M}$ in terms of the spectral idempotents of $U$. We give a short proof along the same lines for completeness. 
\begin{lem}\label{mhatidem}
	Suppose that $U$ is the transition matrix of the Szegedy quantum walk of an arbitrary Markov chain, and suppose that $U$ has spectral decomposition $U=\sum_{r=1}^m\lambda_r F_r$. Then
\[ \widehat{M}_{xy}=  \sum_{r=1}^m \sum_{z \in \Omega} (\bold{e}_x \otimes \bold{e}_z)^* (F_rS) \circ \overline{(F_rS)} \bold{e}_y. \]
\end{lem}

\begin{proof}
	Let $v = S \bold{e}_y$. Using the spectral decomposition of $U$, we can express the time averaged distribution at time $T$ as
\begin{eqnarray*}
\begin{split}
	\frac{1}{T}\sum_{t=0}^{T-1} \big(U^t v \circ \overline{U^tv}\big) &= \frac{1}{T}\sum_{t=0}^{T-1} \Big(\sum_r \lambda_r^t F_rv \Big) \circ \Big(\overline{\sum_s \lambda_s^t F_s v}\Big) \\
	&= \sum_r \big(F_rv \circ \overline {F_r v}\big) + \sum_{r \neq s} \Big( \frac{1}{T}\sum_{t=0}^{T-1} (\lambda_r \overline{\lambda_s})^t \Big) \big(F_rv \circ \overline{F_s v}\big)  \\
	&= \sum_r \big(F_rv \circ \overline {F_r v}\big) + \frac{1}{T} \sum_{r \neq s} \Big(\frac{1-(\lambda_r \overline{\lambda_s})^T}{1-\lambda_r \overline{\lambda_s}} \Big) \big(F_rv \circ \overline{F_s v}\big).  \\
\end{split}
\end{eqnarray*}
Now observe that since each $\lambda_r$ has modulus equal to one then for $r \neq s$ we have
\begin{eqnarray*}
	\Big|\frac{1-(\lambda_r \overline{\lambda_s})^T}{1-\lambda_r \overline{\lambda_s}}\Big| \leq \frac{2}{|1-\lambda_r \overline{\lambda_s}|}
\end{eqnarray*}
hence the sum 
\begin{eqnarray*}
	\frac{1}{T} \sum_{r \neq s} \Big(\frac{1-(\lambda_r \overline{\lambda_s})^T}{1-\lambda_r \overline{\lambda_s}} \Big) \big(F_rv \circ \overline{F_s v}\big)
\end{eqnarray*}
approaches $0$ as $T$ approaches $\infty$. It therefore follows that
\begin{eqnarray*}
	\lim_{T \rightarrow \infty}\frac{1}{T}\sum_{t=0}^{T-1} \big(U^t v \circ \overline{U^tv}\big) = \sum_r \big(F_rv \circ \overline {F_r v}\big),
\end{eqnarray*}
which implies the claimed identity. 
\end{proof}

We now prove a more precise identity for $\widehat{M}$ using the spectral decomposition of (\ref{specdecomp}) and Lemma~\ref{mhatidem}. For a matrix $M$ we let $M^{\circ 2}$ denote the matrix obtained from $M$ by squaring its entries.

\begin{thm}\label{mhat}
Let $P$ be the transition matrix of an irreducible, aperiodic, reversible Markov chain. Then the average mixing matrix of the Szegedy quantum walk of $P$ satisfies
\begin{eqnarray}\label{Midentity}
	\widehat{M}= \sum_{r=1}^m E_r^{\circ 2} - \frac{1}{2}(I - P^T)\sum_{r=2}^m\frac{1}{1-\lambda_r^2}E_r^{\circ 2}.
\end{eqnarray}
\end{thm}

\begin{proof}
Since $P$ is assumed to be reversible then $P$ is similar to $D$ and hence the matrices have the same spectrum. Moreover, as $P$ is irreducible and aperiodic then $-1$ is not an eigenvalue of $P$ and thus $-1$ is not an eigenvalue of $D$. Therefore from (\ref{specdecomp}) and writing each eigenvalue $\lambda_r$ of $D$ as $\lambda_r = \cos(\theta_r)$ the spectral decomposition of $U$ restricted to $\col(S)$ is
\begin{eqnarray*}
	1 \cdot SE_1S^* + \sum_{r=2}^m \Big( e^{i \theta_r} F_r^+ + e^{-i\theta_r}F_r^-\Big).
\end{eqnarray*}
We first directly compute $F_r^+ S\bold{e}_y$ for $r \geq 2$ and $y \in \Omega$:
\begin{eqnarray*}
	\begin{split}
		F_r^+ S\bold{e}_y 
		&= 
		\frac{1}{2\sin^2(\theta_r)}\Big(S-e^{i\theta_r}RS\Big)E_r\Big(S-e^{i\theta_r}RS\Big)^* S\bold{e}_y
		\\
		&= 
		\frac{1}{2\sin^2(\theta_r)}\Big(SE_r  -e^{i\theta_r}RSE_r  -\lambda_r e^{-i\theta_r}SE_r  + \lambda_r RSE_r \Big)\bold{e}_y,
		\\
	\end{split}
\end{eqnarray*}
where we have applied the identities $S^*RS = D$ and $DE_r = \lambda_r E_r$ for all $r$.
The entry of this vector labeled by the arc $(x,z)$ is then 
\begin{eqnarray*}
	\begin{split}
		\frac{1}{2\sin^2(\theta_r)}
		\Big(
		\sqrt{p(x,z)} (E_r)_{xy} 
		-e^{i\theta_r}\sqrt{p(z,x)} (E_r)_{zy} \\
		-\lambda_r e^{-i\theta_r}\sqrt{p(x,z)} (E_r)_{xy} 
		+ \lambda_r \sqrt{p(z,x)} (E_r)_{zy}
		\Big).  
		\\
	\end{split}
\end{eqnarray*}
The real part of this entry simplifies to $\frac{1}{2}\sqrt{p(x,z)} (E_r)_{xy}$, and the imaginary part is 
\begin{eqnarray*}
	\frac{1}{2\sin(\theta_r)}\Big( \lambda_r \sqrt{p(x,z)} (E_r)_{xy}-\sqrt{p(z,x)} (E_r)_{zy} \Big).
\end{eqnarray*}
Together, we deduce that $|(\bold{e}_x \otimes \bold{e}_z)^* F_r^+ S\bold{e}_y|^2$ is given by
\begin{eqnarray*}
\frac{1}{4}p(x,z) (E_r)_{xy}^2+ \frac{1}{4\sin^2(\theta_r)}\Big( \lambda_r \sqrt{p(x,z)} (E_r)_{xy}-\sqrt{p(z,x)} (E_r)_{zy} 
		\Big)^2,
\end{eqnarray*}
or equivalently,
\begin{eqnarray*}
	\frac{1}{4\sin^2(\theta_r)}
		\Big( 
		p(x,z) (E_r)_{xy}^2
		-2\lambda_r D_{xz} (E_r)_{xy} (E_r)_{zy}
		+p(z,x) (E_r)_{zy}^2 
		\Big).
\end{eqnarray*}
Since $F_r^-$ is the complex conjugate of $F_r^+$ then $|(\bold{e}_x \otimes \bold{e}_z)^* F_r^+ S\bold{e}_y|^2$ and $|(\bold{e}_x \otimes \bold{e}_z)^* F_r^- S\bold{e}_y|^2$ are equal. Applying Lemma~\ref{mhatidem} the $(x,y)$ entry of $\widehat{M}$ is then the sum of $\sum_{z \in \Omega} p(x,z)(E_1)_{xy}^2$ and 
\begin{eqnarray*}
	\sum_{z \in \Omega} \sum_{r = 2}^m \frac{1}{2\sin^2(\theta_r)}\Big(p(x,z)(E_r)_{xy}^2 - 2\lambda_r D_{xz} (E_r)_{xy} (E_r)_{zy} + p(z,x) (E_r)_{zy}^2 \Big). \\
\end{eqnarray*}
Applying the fact that $P$ is row-stochastic, and that $\sum_{z \in \Omega} D_{xz}(E_r)_{zy}$ is the $(x,y)$-entry of the product $DE_r = \lambda_r E_r$, this becomes
\begin{eqnarray*}
	\begin{split}
		\widehat{M}_{xy} &= (E_1)_{xy}^2 + \sum_{r = 2}^m \frac{1}{2\sin^2(\theta_r)} \Big( (E_r)_{xy}^2 - 2\lambda_r^2 (E_r)_{xy}^2 + (P^TE_r^{\circ 2})_{xy} \Big), \\
		&= \sum_{r=1}^m(E_r^{\circ 2})_{xy} + \frac{1}{2}\sum_{r=2}^m \frac{1}{1-\lambda_r^2} \Big((P^TE_r^{\circ 2})_{xy} - (E_r^{\circ 2})_{xy} \Big). \\
	\end{split}
\end{eqnarray*}
Writing this as a matrix gives the result.
\end{proof}

\section{Continuous Quantum Walks}\label{continuouswalks}

For any Hermitian matrix, $H$, the matrix exponential 
\begin{eqnarray*}
	\exp(itH) := \sum_{k=0}^\infty \frac{(it)^k}{k!}H^k
\end{eqnarray*}
is a unitary matrix which defines a continuous quantum walk with respect to $H$. In physics, the matrix $H$ is often referred to as the \textit{Hamiltonian} of the evolution. We define a continuous quantum walk on a Markov chain by the unitary matrix $\exp(itD)$. By unitarity, the columns of the matrix $\exp(itD) \circ \exp(-itD)$ are probability densities, which we can time-average as with the discrete case to obtain the average mixing matrix
\begin{eqnarray}
	\widehat{M}_C : = \lim_{T \rightarrow \infty} \frac{1}{T} \int_0^T \exp(itD) \circ \exp(-itD) \ \textnormal{d}t.
\end{eqnarray}
The subscript in $\widehat{M}_C$ serves to remind the reader that this mixing matrix arises from a continuous quantum walk rather than a discrete one. As was shown in \cite[Theorem 3.2]{R2007} this average mixing matrix can be written in terms of the spectral idempotents of $D$ as
\begin{eqnarray}\label{contmixingidem}
	\widehat{M}_C = \sum_{r=1}^m E_r^{\circ 2}.
\end{eqnarray}
Note that with this notation the identity presented in (\ref{Midentity}) can be expressed as
\begin{eqnarray}\label{Midentity2}
	\widehat{M}= \widehat{M_C} - \frac{1}{2}(I - P^T)\sum_{r=2}^m\frac{1}{1-\lambda_r^2}E_r^{\circ 2}.
\end{eqnarray}
The average mixing matrix $\widehat{M}_C$ has received some attention in the case of a continuous quantum walk on a graph, meaning that the Hamiltonian is taken to be either the adjacency matrix of Laplacian matrix of a graph. On the other hand, there seems to be little known about $\widehat{M}_C$ for reversible Markov chains. We will say that the Szegedy quantum walk $U$ or continuous quantum walk $\exp(itD)$ is \textit{average uniform mixing} if its corresponding average mixing matrix is equal to $\frac{1}{n}J$, where $J$ denotes the all-ones matrix. The following lemma will be useful for our analysis.

\begin{lem}\cite[Lemma 17.1]{G2012}\label{psd}
	If the $n \times n$ matrices $M_1,...,M_k$ are positive semidefinite and $\sum_r M_r$ is a multiple of $J$, then $M_r$ is a multiple of $J$ for each $r$.
\end{lem}

We can now completely characterize average uniform mixing in reversible Markov chains in terms of the eigenvectors and eigenvalues of $D$. The forward direction of our result was proven in the proof of \cite[Lemma 17.2]{G2012} for continuous quantum walks on graphs. We show that the same reasoning holds for reversible Markov chains as well. To state our results, we will say that an eigenvalue is \textit{simple} if it has multiplicity one, and we refer to a vector with all entries equal to $+1$ or $-1$ as a $+1, -1$ vector.

\begin{thm}\label{contmixing}
	The continuous quantum walk $\exp(itD)$ is average uniform mixing if and only if every eigenvalue of $D$ is simple, and every eigenvector of $D$ is a multiple of a $+1, -1$ vector.
\end{thm}

\begin{proof}
Suppose that $\exp(itD)$ is average uniform mixing. Then this means, by definition and (\ref{contmixingidem}), that $\sum_{r=1}^m E_r^{\circ 2} = \frac{1}{n}J$. Since each spectral idempotent $E_r$ is a projection, then its eigenvalues are all nonnegative, so that each $E_r$ is positive semidefinite. By the Schur-product theorem, each $E_r^{\circ 2}$ is also positive semidefinite. Therefore Lemma~\ref{psd} implies that $E_r^{\circ 2}$ is a multiple of $J$ for each $r=1,...,m$. Writing $|(E_r)_{xy}|=\alpha$, then it follows that
\begin{eqnarray*}
	\alpha = |\langle E_r \bold{e}_x, \bold{e}_x \rangle| = |\langle E_r \bold{e}_x, E_r \bold{e}_x \rangle| = n \alpha^2,
\end{eqnarray*}
which then implies that $\alpha = 1/n$. As a consequence, $\tr(E_r) = 1$ and since the trace of $E_r$ is equal to the dimension of the eigenspace it projects onto, then we conclude that each eigenvalue of $P$ is simple. As each $E_r$ is now a rank-one projection and $|(E_r)_{xy}|=\frac{1}{n}$ for all $x, y \in \Omega$, then we deduce that every eigenvector of $D$ must be a multiple of a $+1, -1$ vector.

Conversely, if every eigenvalue of $D$ is simple then each idempotent $E_r$ is a rank-one projection. We can therefore express $E_r$ as $E_r = vv^*$ for some normalized eigenvector $v$ with eigenvalue $\lambda_r$. By hypothesis, $v$ is a multiple of a $+1, -1$ vector, so each of its entries must be equal to $\frac{1}{\sqrt{n}}$ in absolute value. Hence, $|(E_r)_{xy}|=\frac{1}{n}$ for all $x,y \in \Omega$. We compute $\widehat{M}_C$ from (\ref{contmixingidem}) as 
\begin{eqnarray*}
	\widehat{M}_C = \sum_{r=1}^n \frac{1}{n^2}J = \frac{1}{n}J,
\end{eqnarray*}  
so $\exp(itD)$ is indeed average uniform mixing, completing the proof.
\end{proof}

We lastly prove a relationship between average uniform mixing in the continuous quantum walk and Szegedy quantum walk.

\begin{thm}\label{avguniformmix}
Let $P$ be a symmetric, irreducible and aperiodic Markov chain. If the continuous quantum walk $\exp(itP)$ is average uniform mixing, then so is the Szegedy quantum walk $U$.
\end{thm}

\begin{proof}
Suppose that $\widehat{M}_C = \frac{1}{n} J$. Then by Lemma~\ref{psd} each of the matrices $E_r^{\circ 2}$ is equal to a multiple of $J$. Since $P$ is symmetric it has uniform stationary distribution, hence $JP = J$, or equivalently, $(I-P^T)J=0$. From this it follows that $(I - P)E_r^{\circ 2} = 0$ for all $r=1,...m$, so Theorem~\ref{mhat} implies $\widehat{M} = \frac{1}{n}J = \widehat{M}_C$, as claimed.
\end{proof}

\section{Properties of $\widehat{M}$}\label{properties}

We list some of the basic properties of the matrix $\widehat{M}$ defined in Section \ref{average}. To state our results, we recall that an \textit{automorphism} of a Markov chain is a bijection $\sigma : \Omega \rightarrow \Omega$ such that $p(\sigma(x), \sigma(y)) = p(x,y)$ for all $x, y \in \Omega$. 

\begin{thm}\label{mhatproperties}
	The average mixing matrix of the Szegedy quantum walk satisfies
\begin{enumerate}[(a)]
	\item If $P$ is symmetric and has entries in a field $F$ such that $\Q \subseteq F \subseteq \R$, then so does $\widehat{M}$.
	\item If $\sigma$ is any automorphism of $P$, then $\widehat{M}_{xy} = \widehat{M}_{\sigma(x), \sigma(y)}$ for all $x, y \in \Omega$.
	\item $\widehat{M}$ is column-stochastic. 
\end{enumerate}
\end{thm}

\begin{proof}
\begin{enumerate}[(a)]  
	\item We generalize the argument presented in \cite[Lemma 10.7]{GZ2019}. Note that since $P$ is real symmetric, then $P = D$. Let $K$ be a splitting field for the characteristic polynomial of $P$, which therefore defines an algebraic extension of $F$. Since $F$ is contained in $\R$, then $F$ has characteristic zero. An algebraic extension of a field of characteristic zero is separable, hence $K$ is a separable extension of $F$. As $K$ is separable over $F$ and is the splitting field of a polynomial then the field extension $K/F$ is Galois. Next consider a Galois automorphism $\tau \in \Gal(K/F)$. Since $\tau$ fixes the field $F$ then $P$ may be expressed as
\begin{eqnarray*}
	P = \tau(P) = \sum_{r=1}^m \tau(\lambda_r) \tau(E_r).
\end{eqnarray*}
	As $\tau$ is an automorphism, the matrices $\tau(E_1),...,\tau(E_m)$ must be orthogonal and idempotent. Additionally, every Galois automorphism acts as a permutation of the eigenvalues of $P$, hence by the uniqueness of the spectral decomposition we deduce that $\tau$ also acts as a permutation of the set of spectral idempotents $\{E_r\}_{r=1}^m$. As a consequence, the matrix $\sum_{r=1}^m E_r^{\circ 2}$ is fixed by every element of $\Gal(K/F)$ and therefore must have entries in $F$. Continuing, since $\tau$ fixes the field $F$ then $\tau$ fixes the entries of the matrix $\frac{1}{2}(I - P)$. Since $\tau$ acts as a permutation on the eigenvalues of $P$, and fixes $\Q$, then $\tau(\lambda_r) \neq 1$ for any $r \geq 2$. Therefore $\tau$ acts as a permutation on the set $\{ \frac{1}{1-\lambda_r^2}E_r^{\circ 2} : r =2,...,m\}$, which implies that the sum $\sum_{r=2}^m \frac{1}{1-\lambda_r^2}E_r^{\circ 2}$ is fixed by $\tau$. Applying Theorem~\ref{mhat} we see that $\widehat{M}$ is fixed by every automorphism in $\Gal(K/F)$ and thus must have entries in $F$. 
	\item Suppose that $\sigma$ is an automorphism of $P$, and let $P_\sigma$ denote its permutation representation, meaning that $P_\sigma\bold{e}_x =\bold{e}_{\sigma(x)}$ for all $x \in \Omega$. A simple computation shows that $R$ commutes with $P_\sigma \otimes P_\sigma$, and $SP_\sigma = (P_\sigma \otimes P_\sigma)S$. From this we deduce that $USP_\sigma \bold{e}_y = (P_\sigma \otimes P_\sigma) U S \bold{e}_y$, which then implies that
\begin{eqnarray}\label{autom}
	\frac{1}{T}\sum_{t=0}^{T-1} (U^tS) \circ (\overline{U^tS}) P_\sigma \bold{e}_y = (P_\sigma \otimes P_\sigma) \frac{1}{T}\sum_{t=0}^{T-1} (U^tS) \circ (\overline{U^tS})\bold{e}_y.
\end{eqnarray}
Since $P_\sigma$ is a permutation matrix then the all ones vector is both a left and right eigenvector with eigenvalue $1$, so that $\sum_{z \in \Omega} (\bold{e}_x \otimes \bold{e}_z)^* (P_\sigma \otimes P_\sigma) = \sum_{z \in \Omega} (P_\sigma^T \bold{e}_x \otimes \bold{e}_z)^T$. Together with equation (\ref{autom}) we obtain
\begin{eqnarray}\label{autom2}
\begin{split}
	\sum_{z \in \Omega} (\bold{e}_{\sigma(x)} \otimes \bold{e}_z)^* \frac{1}{T}\sum_{t=0}^{T-1} (U^tS) \circ (\overline{U^tS}) \bold{e}_{\sigma(y)} \\
	= \sum_{z \in \Omega} (\bold{e}_x \otimes \bold{e}_z)^* \frac{1}{T}\sum_{t=0}^{T-1} (U^tS) \circ (\overline{U^tS})  \bold{e}_y.
\end{split}
\end{eqnarray}
Taking the limit of both sides of (\ref{autom2}) as $T$ approaches $\infty$ we conclude that $\widehat{M}_{xy} = \widehat{M}_{\sigma(x), \sigma(y)}$, as claimed.
	\item To show that $\widehat{M}$ is column-stochastic we will show that $\bold{1} \widehat{M} = \bold{1}$, where $\bold{1}$ denotes the all-ones row vector. As $P$ is row-stochastic then $\bold{1}P^T = \bold{1}$, so applying Theorem~\ref{mhat} implies that 
\begin{eqnarray}
	\bold{1}\widehat{M} = \bold{1} \sum_{r=1}^m E_r^{\circ 2}.
\end{eqnarray}
We can then evaluate the matrix equation $\bold{1}E_r^{\circ 2}$ directly,
\begin{eqnarray}
	\begin{split}
		(\bold{1}E_r^{\circ 2})_y &= \sum_{x \in \Omega} (E_r)_{xy}^2 \\
		&= \langle E_r \bold{e}_y, E_r \bold{e}_y \rangle \\
		&= \langle \bold{e}_y, E_r^2 \bold{e}_y \rangle \\
		&= \langle \bold{e}_y, E_r \bold{e}_y \rangle \\
		&= (E_r)_{yy}, \\
	\end{split}
\end{eqnarray}
where we have used the fact that the matrices $E_r$ are symmetric and idempotent. Since $\sum_{r=1}^m E_r = I$ then we conclude that $\bold{1} \widehat{M} = \bold{1} \sum_{r=1}^m E_r^{\circ 2} = \bold{1}$ as claimed. 
\end{enumerate}
\end{proof}


\section{Examples of Average Uniform Mixing}\label{uniformmixingexamples}

In this section we will construct an infinite family of Markov chains with arbitrarily large size whose continuous quantum walk and Szegedy quantum walk are average uniform mixing. To build up to this construction we begin by considering the smallest non-trivial symmetric Markov chain. 

Let $P$ be the transition matrix defined by 
\begin{eqnarray}\label{twostatechain}
	P := \left(\begin{array}{cc}
		p & 1-p \\
		1-p & p \\
	\end{array}\right),
\end{eqnarray}
for any $p \in (0,1)$, which defines a symmetric Markov chain. We determine its spectral decomposition as follows. Since $P$ is symmetric then the all-ones vector is an eigenvector with eigenvalue $1$, and by considering the trace of $P$ the remaining eigenvalue of $P$ must be $2p - 1$. The projection onto the eigenspace corresponding to the eigenvalue $1$ is thus $E_0:=\frac{1}{2}J$, and since the spectral idempotents sum to the identity then the remaining idempotent is $E_1:=I - \frac{1}{2}J$. Therefore the spectral decomposition of $P$ is
\begin{eqnarray}\label{twostatespecdecomp}
	P = 1 \cdot E_0 + (2p-1) \cdot E_1.	
\end{eqnarray}

\begin{thm}\label{twostateavgmixing}
	For every $p \in (0,1)$, both the continuous quantum walk and Szegedy quantum walk on the symmetric Markov chain defined by (\ref{twostatechain}) are average uniform mixing.
\end{thm}

\begin{proof}
	From our description of the spectral decomposition of $P$ given in (\ref{twostatespecdecomp}) the matrix $P$ has two simple eigenvalues, and every eigenvector is either a multiple of the all-ones vector, or the vector $(1 \ - 1)^T$. The result now follows from Theorem~\ref{contmixing} and Theorem~\ref{avguniformmix}. 
\end{proof}

We will now build on the example given in Theorem~\ref{twostateavgmixing} to construct an infinite family of Markov chains with arbitrarily large size which are average uniform mixing. Our construction will be based on the tensor product of two Markov chains. If $P$ and $Q$ are transition matrices for any two symmetric Markov chains, then $P \otimes Q$ is both symmetric and doubly-stochastic, hence is the transition matrix for a larger symmetric Markov chain. Moreover, the eigenvectors and eigenvalues of $P \otimes Q$ are determined by those of $P$ and $Q$. If $v_1,...,v_n$ and $w_1,...,w_m$ are orthonormal bases of eigenvectors for $P$ and $Q$, respectively, with respective eigenvalues $\lambda_1,...,\lambda_n$ and $\theta_1,...,\theta_m$, then $v_i \otimes w_j$ is an eigenvector of $P \otimes Q$ with eigenvalue $\lambda_i \theta_j$ for all $i=1,...,n$ and $j=1,...,m$. Additionally, the set of eigenvectors $\{ v_i \otimes w_j : i = 1,...,n, j=1,...,m \}$ forms an orthonormal basis.

\begin{thm}\label{tensormixing}
	Let $P_1,...,P_k$ be symmetric Markov chains, and suppose that for each $i=1,...,k$ the continuous quantum walk on $P_i$ is average uniform mixing. If the product $\prod_{j=1}^k \lambda(j)$ is distinct for each choice of eigenvalues $\lambda(1) \in \Spec(P_1), \lambda(2) \in \Spec(P_2),...,\lambda(k) \in \Spec(P_k)$, then the continuous quantum walk on the tensor product $P_1 \otimes ... \otimes P_k$ is average uniform mixing. 
\end{thm}

\begin{proof}
	From Theorem~\ref{contmixing}, since each $P_i$ is average uniform mixing then each matrix $P_i$ has simple eigenvalues. Since the eigenvalues of the tensor product $P_1 \otimes P_2 \otimes ... \otimes P_k$ are $\prod_{j=1}^k \lambda(j)$ for $\lambda(1) \in \Spec(P_1), \lambda(2) \in \Spec(P_2),...,\lambda(k) \in \Spec(P_k)$, then by our hypothesis on these products, the eigenvalues of $P_1 \otimes P_2 \otimes ... \otimes P_k$ are all simple. From our description above on the eigenvectors of a tensor product, the eigenvectors of $P_1 \otimes ... \otimes P_k$ are all of the form $v_1 \otimes v_2 \otimes ... \otimes v_k$, where $v_i$ is an eigenvector of $P_i$. By Theorem~\ref{contmixing} each eigenvector $v_i$ is a multiple of a $+1, -1$ vector, hence $v_1 \otimes v_2 \otimes ... \otimes v_k$ is a multiple of a $+1, -1$ vector. The Markov chain $P_1 \otimes ... \otimes P_k$ is therefore average uniform mixing by Theorem~\ref{contmixing}.
\end{proof}

We now present our infinite family of Markov chains with the desired property.

\begin{thm}\label{uniformmixingchain}
	Let $q_1,...,q_k$ be any distinct $k$ odd primes, and define the symmetric Markov chains $P_1,...,P_k$ by 
\begin{eqnarray*}
	P_i = \left(\begin{array}{cc}
		p_i & 1-p_i \\
		1-p_i & p_i \\
	\end{array}\right)
\end{eqnarray*}
where $p_i := \frac{1}{2}(1+\frac{1}{q_i})$. Then the both the continuous quantum walk and Szegedy quantum walk on the symmetric Markov chain $P_1 \otimes P_2 \otimes ... \otimes P_k$ are average uniform mixing.
\end{thm}

\begin{proof}
	By our choice of $p_i$, the eigenvalues of $P_i$ are $1$ and $\lambda(i) := \frac{1}{q_i}$. The eigenvalues of $P_1 \otimes ... \otimes P_k$ are then given by the products
\begin{eqnarray*}
	\lambda_I:= \prod_{i \in I} \lambda(i)
\end{eqnarray*}
for $I \subseteq \{1,...,k\}$. The equality $\lambda_I = \lambda_{I^\prime}$ for $I, I^\prime \subseteq \{1,...,k\}$ 
is equivalent to the equality
\begin{eqnarray*}
	\prod_{i \in I} q_i = \prod_{j \in I^\prime} q_j,
\end{eqnarray*}
and since the $q_1,...,q_k$ are all prime this holds if and only if $I = I^\prime$. As a result the eigenvalues of $P_1 \otimes ... \otimes P_k$ are all simple, and so the result follows by Theorem~\ref{tensormixing} and Theorem~\ref{avguniformmix}.
\end{proof}

We note that the construction above also holds in exactly the same way if we define $p_i=\frac{1}{2}(1-\frac{1}{q_i})$ for $i=1,...,k$.

\section{Discussion and Open Questions}

Understanding the limiting behavior of the Szegedy quantum walk is important for determining its potential role in quantum sampling algorithms. Our contribution to this is the study of an average mixing matrix, $\widehat{M}$, which encodes the limiting probability distribution of the quantum walk over the vertices given that the initial state is a state which is concentrated on a vertex. One of our main results in Theorem~\ref{mhat} gave a formula for this matrix in terms of the spectral decomposition of $P$ and its discriminant, $D$. Furthermore, in Theorem~\ref{avguniformmix} we showed that average uniform mixing in the continuous quantum walk implies average uniform mixing in the Szegedy quantum walk. This is an interesting result since simulating a discrete quantum walk is generally easier than simulating a continuous one. We leave it as a problem for future research to determine if the converse holds or to find a counterexample. 

In Theorem~\ref{uniformmixingchain} we constructed a family of symmetric Markov chains with arbitrarily large size which admits average uniform mixing in both its continuous quantum walk and Szegedy quantum walk. Since a symmetric Markov chain converges to the uniform distribution this shows that it is possible to sample from a Markov chain's stationary distribution using a quantum walk. Moreover, in \cite[Lemma 17.2]{G2012} it was shown that no continuous quantum walk on a graph with more than two vertices admits average uniform mixing. Since a symmetric Markov chain can be thought of as a weighted graph, it means that average uniform mixing is possible for some weighted graphs with more than two vertices, but no unweighted graph with more than two vertices. 

For applications in approximate counting there are many different Markov chains we can define to approximately count the same object. This is because in practice the only requirements are that the chain converges to the uniform distribution and respects a certain graph topology, meaning that there is some freedom in choosing the precise transition probabilities. It would be interesting to determine if it is always possible to define a Markov chain on an arbitrary digraph such that its quantum walk is average uniform mixing.


\section*{Acknowledgements}

The author would like to thank Steven Herbert and Pablo Andres-Martinez for their helpful comments on an early draft of this paper.



\bibliographystyle{elsarticle-num} 
\bibliography{bibliography.bib}





\end{document}